# Ionospheric response to the space weather event of 18 November 2003 - An investigation

Pankaj Kumar[1,$], Wahab Uddin[1], Alok Taori[2], Ramesh Chandra[3] & Shuchi Bisht[4]

[1]Aryabhatta Research Institute of Observational Sciences (ARIES), Nainital 263 129, Uttarakhand, India
[2]National Atmospheric Research Laboratory (NARL), Gadanki 517 112, A P, India
[3]Observatoire de Paris, LESIA, UMR8109 (CNRS), F- 92195, Meudon Principal Cedex, France
[4]Department of Physics, DSB Campus, Kumaun University, Nainital 263 002
[$]E-mail: pkumar@aries.res.in

The present study explores the ionospheric effects of the well cited solar flare events (M3.2, M3.9/2N) of 18 November 2003 associated with CMEs. The Hα observations of these flares (taken with 15 cm Solar Tower Telescope at ARIES, Nainital) have been analysed to see an association of these flare events with the geomagnetic storm occurred on 20 November 2003. The ionospheric data from Puerto Rico (18.5°N, 67.2°W), Dyess (32.4°N, 99.7°W) and Millstone Hill (42.6°N, 71.5°W) together with the disturbance storm time indices (Dst index) variability exhibited a corresponding associations having a delay in the solar wind parameters, triggered by these flare events.

## 1 Introduction

The ionospheric response to the solar events is an integral part of space weather. Solar flares are important to space weather mainly because of their connection to the coronal mass ejection (CME) with often being the seed for the CMEs because they have an important role in solar wind particle acceleration. The enhanced X-ray and extreme ultraviolet (EUV) solar radiation during a flare causes a dramatic increase in ionospheric ionization, with several consequences for radio propagation and telecommunication systems[1,2]. Several investigators also reported the significant variations in the low latitude ionosphere-thermosphere behaviour[3] compared to the quiet time[4].

When the CME hits the earth's magnetosphere, geomagnetic storms are triggered. The interplanetary magnetic field (IMF), carried by solar wind, has three components and when the z component of IMF ($B_Z$) becomes southward to the geomagnetic field lines, magnetic reconnection takes place, which is the most suitable condition for the transfer of solar wind momentum and energy to the magnetosphere. Owing to the gradient and curvature drifts acting on the charged particles in the magnetosphere, the ring current flows in the westward direction around the earth at radial distance 2-7 $R_e$, having geo-magnetically trapped 10-200 keV ions (mainly $H^+$, $He^+$ and $O^+$) and electrons. The magnetic field associated with the ring current is in the same direction to that of geomagnetic field, thus an increase in the ring current causes a decrease in the geomagnetic field. The variation in the geomagnetic field is called geomagnetic storms.

The Dst index (ring current index) is an indicative of the total energy content of the particles responsible for ring current. The geomagnetic storms with Dst between –200 and -100 nT are classified as 'intense' and other events with Dst < -200 nT as 'super-intense'. Super-storms take place with Dst < -300 nT as observed during October-November 2003. These geomagnetic storms affect the earth's ionosphere. In the present paper, the ionospheric effect of geomagnetic storms associated with solar flare events on 18 November 2003 at different latitudes have been presented by taking the ionosonde data from Puerto

Rico (18.5°N, 67.2°W), Dyess (32.4°N, 99.7°W) and Millstone Hill (42.6°N, 71.5°W).

## 2 Solar flare events and CMEs

The period of October-November 2003 was characterized as the highest levels of solar activity to date when solar cycle 23 was well into a declining phase. During this period, on 18 November 2003 three solar flares (C3.8, M3.2 and M3.9) were observed from the active region NOAA 10501 with a βγδ type configuration. The C3.8 is the impulsive flare and M3.2 and M3.9/2N are the homologous flares. The M3.2 and M3.9 are the good examples of the solar flares characterized by two-ribbon long duration events. These flares are associated with filament eruptions and two CMEs which are responsible for strong geomagnetic storms. The evolution of flares and filament eruption are shown in Fig. 1. Using the multi-wavelength data, detail study of these flares is given by Chandra[5].

Large Angle and Spectrometric Coronagraph (LASCO) and SOHO Extreme ultraviolet Imaging Telescope (EIT) observed two full halo CMEs (CME I and CME II) associated with these flares (Fig. 2). A wide faint loop front of CME I was seen in C2 at 08:06 hrs UT filling the south-east quadrant but brightest in the S at 08:26 hrs UT, this front had faint extensions up to the North pole. At 08:50 hrs UT, a second, much brighter, front of CME II appeared spanning ~160 deg from the SE to NW and with fainter extensions to the N pole. Thus, between the two sets of fronts, there was full coverage of the C2 occulter by 08:50 hrs UT, albeit rather faint in the NE. Both CMEs were very fast having speed 1222 and 1660 km s$^{-1}$ with PA 169 and 206, respectively.

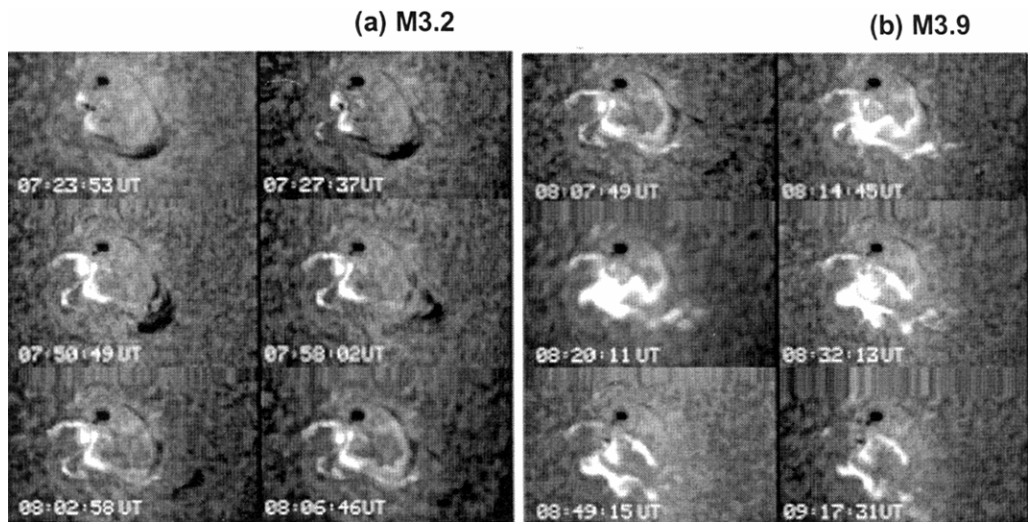

Fig. 1—Hα observations of M3.2 and M3.9 solar flares on 18 November 2003 taken with 15 cm Coude solar tower telescope at ARIES, Nainital, India. The FOV of each image is 574" × 384". The North is up and west is to the right

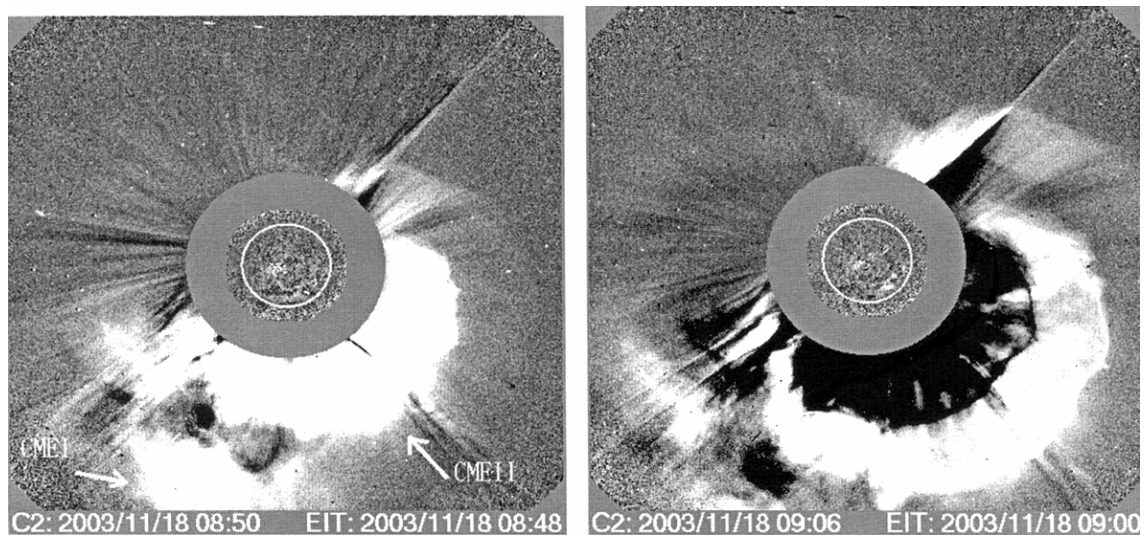

Fig. 2—Two full halo CMEs associated with the solar flare observed by SOHO LASCO C2

These CMEs were associated with complex activity in and around AR 10501 during 07:36 - 09:00 hrs UT. Two flares were observed between 07:36 - 08:00 hrs UT and 08:12 - 09:00 hrs UT in Hα, which were also observed by SOHO/EIT. The first flare was centered at N03E18 and the second a little to the south and west. Additionally, a filament channel to the southwest of the active region was activated following the first flare and subsequently erupted. Both an EIT wave and dimming were observed in association with these events[6].

During flare events M3.2 and M3.9, the Hα filament erupted in the southwest direction. To see the temporal correlation between the filament eruptions and CME (here CME II), the temporal evolution of filament eruption and CME II have been plotted in Fig. 3. The height-time plot of the filament eruption and CME II has been compared, because both the filament and CME II are erupting in the same direction i.e. southwest. The total energy content (kinetic, potential and magnetic) of the CME II and filament to be associated with the solar flare has been estimated. The estimated values of the energy for the CME II and filament are ~9.6×10$^{31}$ and ~1.7×10$^{32}$ ergs, respectively. From the comparison of filament and CME II energy, it is clear that the filament energy is sufficient to produce this CME. This confirms an association of CME II with the filament eruptions.

**3 Geomagnetic variations**

The CMEs of 18 November 2003 were associated with the M3.2 and M3.9 flares, which were closely related in space and time through a chain of complex activity process in active region 10501. The CME near the Sun had a sky-plane speed of ~1660 km s$^{-1}$, but the associated magnetic cloud (MC) arrived with a speed of only 730 km s$^{-1}$.

The full halo CMEs associated with these flares took 47.5 h to travel from sun to earth. The arrival, around 07:45 hrs UT on 20 November, of the shock front of the powerful halo CME triggered a severe geomagnetic storm[7]. This was the largest geomagnetic storm of solar cycle 23. In ACE and SOHO/CELIAS data, the signature of this arrival was clearly visible. The solar wind speed jumped suddenly from 430 to 750 km s$^{-1}$. The solar wind flow (ram) pressure is found to have maximum value 16.26 nPa at 09:30 hrs UT on 20 November.

The z component of IMF ($B_Z$) remains southward up to 13.5 ($T_{B_Z}$) h with a maximum strength of –45.58 nT at 16:00 hrs UT on 20 November. This also might be one of the reasons to produce geomagnetic storm of such a great strength as southward

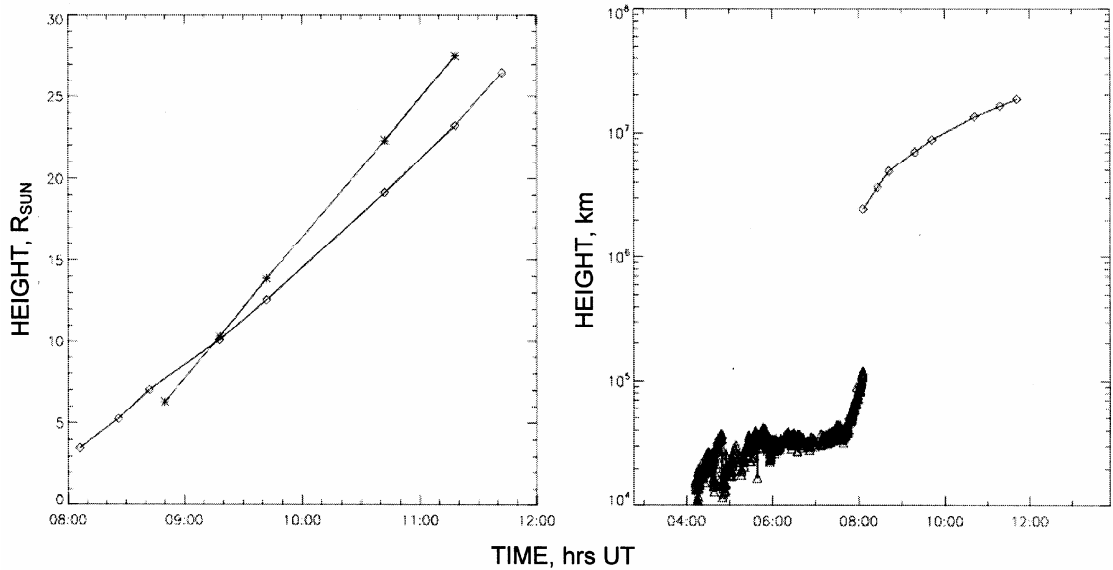

Fig. 3—Height time profile of the CME I(star), CME II (square) (left), and height time plot of CME II (square) along with filament eruption

component of IMF is supportive for magnetic reconnection at the boundary of magnetosphere[8].

Figure 4 shows the plot of Dst index having minimum value of -422 nT at 20:00 hrs UT on 20 November followed by an extended recovery phase, due to a coronal mass ejection (CME) from active region 10501.

## 4 Ionospheric effects

It is understood that geomagnetic storms affect the global circulations in the ionosphere and accordingly variability could be noted at different latitudes. At high latitudes, there is depletion in electron density of F2 layer, which is called negative storm effect. Negative storm effects are quite explainable. But at middle and low latitudes, the nature of ionospheric effects of geomagnetic storms is poorly understood. There may be enhancement in electron density (positive storm effect) or depletion in electron density (negative storm effect). However, these effects are not well understood owing to the lack of comprehensive data[9].

During November 2003 space weather event, significant variations (~15-20%) were noted in ionospheric parameters at different latitudes. Figure 5 shows the ionosonde data for $foF_2$ (critical frequency of $F_2$ layer) and $hmF_2$ (peak height of $F_2$ layer) data of three different latitudes, i.e. Puerto Rico (18.5°N, 67.2°W), Dyess (32.4°N, 99.7°W) and Millstone Hill (42.6°N, 71.5°W) during this event, which shows the noon-time and mid-night $foF2$ and $hmF2$ together with the rate of change of Dst (to see the variability in energy pumped into the magnetosphere). A cross correlation analysis revealed a positive correlation (> 0.2) with a time delay between rate of change of Dst and foF2.

## 5 Results and Discussion

It is clear from the observations and subsequent calculations that the flare events M3.2 and M3.9 triggered enormous CMEs resulting into significant perturbations in the solar-terrestrial environment (such as solar wind speed and particle density data). After the CME arrival at near earth space, first event was the onset of a geomagnetic storm as evident in the magnetospheric ring current index (Dst). The Dst index went down to ~-422 nT. To see the consequences at ionospheric altitudes, the ionospheric data collected from Puerto Rico (18.5°N, 67.2°W), Dyess (32.4°N, 99.7°W), and Millstone Hill (42.6°N,

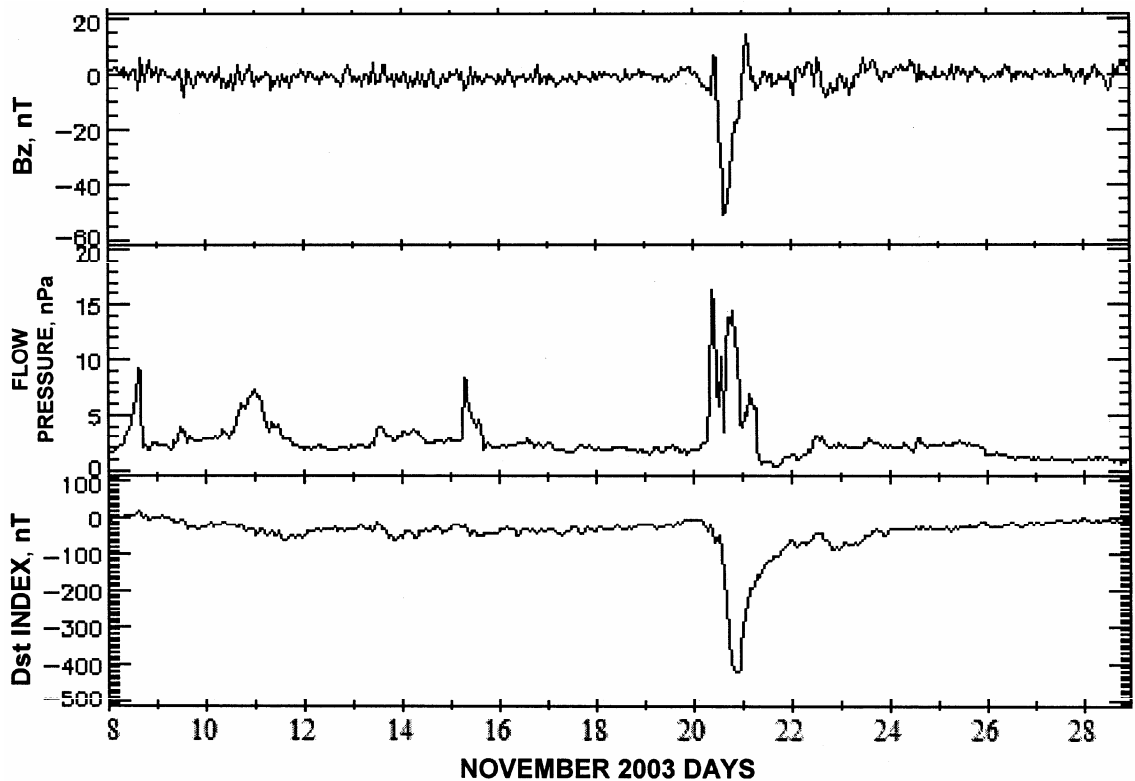

Fig. 4—$B_z$ component of IMF, solar wind flow pressure, and ring current variability during 8-28 November 2003

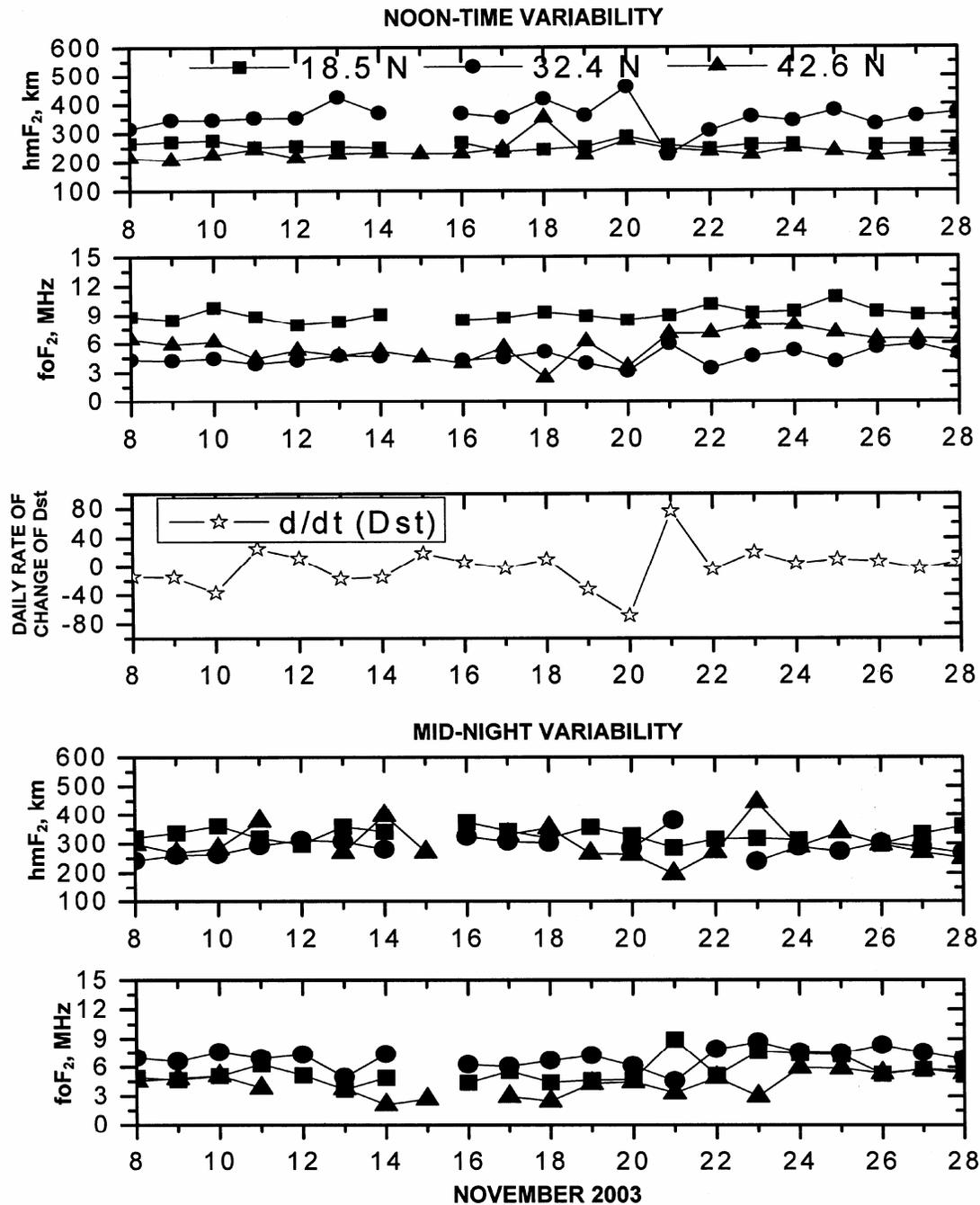

Fig. 5—Noon time and mid night variabilities in foF2 and hmF2 with the rate of change of Dst index at the three different latitudes during 8-28 November 2003

71.5°W) were scrutinized, which show significant perturbations. The perturbation amplitudes during the event varied from high to low latitudes. Through ionospheric data, it was found that mid and high latitudes (42.6°N and 32.4°N) exhibited larger amplitudes of perturbations in response to the geomagnetic disturbances than low latitudes (18.5°N) on the November 2003 event. A correlation analysis between rate of change of Dst and $foF_2$ exhibited the correlation coefficient to peak with a delay for 32.4°N (~5 h) and 18.5°N (~12 h). On the other hand, 42.6°N exhibited no significant correlations indicating that the process of the energy percolation for 42.6°N latitudes is somewhat different from that of 32.4°N and 18.5°N latitudes. The result of this study fits well with the disturbance dynamo theory, which was well examined with the help of DE-2 satellite data by Pant & Sridharan[10] even including the time delayed response of low latitudes.

In the light of above, it seems that the variability observed at low latitude ionosphere might be due to the neutral winds or propagating neutral atmospheric disturbances such as long period gravity wave and tides seeded by the auroral heating, and not by traveling ionospheric disturbances (TIDs) because the disturbance seems to travel towards the equator. From the delay time observed at the two different latitudes (Puerto Rico and Dyess), the velocity of the propagating wave has been calculated which comes out nearly 142 km h$^{-1}$, which is not so high as observed for the TIDs (which often is ~300-600 km h$^{-1}$).

In a nutshell, present report is only exploratory in nature to study the integrated features from sun to the earth's ionosphere or space weather processes as a whole. More data and coordinated studies on space weather events are required to see the variability in the ionospheric responses for a proper characterization of an event.

**Acknowledgements**
The authors wish to acknowledge the ionosonde data used here for 18 November 2003, which was downloaded from the website of Space Physics Data Analysis Resources (SPIDR). The solar wind data is taken from home page of Advanced Composition Explorer (ACE) satellite. The research at ARIES, Nainital is supported by Department of Science and Technology (DST), Government of India.